\documentclass[prd,aps,twocolumn,nofootinbib,preprintnumbers,superscriptaddress,preprintnumbers,balancelastpage,longbibliography]{revtex4-2}
\usepackage[utf8]{inputenc}
\usepackage[colorlinks=true,citecolor=blue,linkcolor=blue]{hyperref}
\usepackage[normalem]{ulem}
\usepackage{amsmath,amssymb, mathrsfs,xfrac}
\usepackage{epsfig}
\usepackage{graphicx}               
\usepackage{color}
\usepackage{slashed}
\usepackage{multirow}
\usepackage{placeins}
\usepackage[dvipsnames]{xcolor}
\usepackage{epstopdf}
\usepackage{soul}
\usepackage{tikz}
\usepackage[capitalise, english]{cleveref}
\usepackage{siunitx}
\usepackage{xspace}
\usepackage{booktabs}
\usetikzlibrary{trees}
\usetikzlibrary{decorations.pathmorphing}
\usetikzlibrary{decorations.markings}
\usepackage{aas_macros}
\usepackage{lettrine}

\newcommand\myshade{80}
\colorlet{mylinkcolor}{ForestGreen}
\colorlet{mycitecolor}{Aquamarine}
\colorlet{myurlcolor}{violet}

\hypersetup{
  linkcolor  = mylinkcolor!\myshade!black,
  citecolor  = mycitecolor!\myshade!black,
  urlcolor   = myurlcolor!\myshade!black,
  colorlinks = true
}

\definecolor{jblue}{RGB}{20,50,100}
\definecolor{npurple}{RGB} {153, 51, 204}
\definecolor{wred}{RGB}{217,0,56}
\definecolor{white}{RGB}{255,255,255}

\definecolor{korange}{RGB}{235, 80,  43}
\definecolor{korange2}{RGB}{245, 100,  63}
\definecolor{kyelloworange}{RGB}{255, 210,  110}
\definecolor{kyelloworange2}{RGB}{240, 170,  90}
\definecolor{kred}{RGB}{204,  102, 153}
\definecolor{kpurple}{RGB}{153,  61, 190}
\definecolor{kpurplelight}{RGB}{213,  161, 230}

 \definecolor{tobycolour}{rgb}{.5,.0,.5}

\DeclareSIUnit\year{yr}
\DeclareSIUnit\pc{pc}
\DeclareSIUnit\ergs{ergs}
\DeclareSIUnit\msun{\ensuremath{M_\odot}}
\allowdisplaybreaks

\newcommand{\eV}{\,\mathrm{eV}}
\newcommand{\surfb}{\,\mathrm{MJy\,sr^{-1}}}

\makeatletter
\providecommand*{\diff}%
  {\command{\lmultau}{\ensuremath{L_\mu-L_\tau}\xspace}
\new@ifnextchar^{\DIfF}{\DIfF^{}}}
\def\DIfF^#1{%
  \mathop{\mathrm{\mathstrut d}}%
    \nolimits^{#1}\gobblespace}
\def\gobblespace{%
  \futurelet\diffarg\opspace}
\def\opspace{%
  \let\DiffSpace\!%
  \ifx\diffarg(%
    \let\DiffSpace\relax
  \else
    \ifx\diffarg[%
      \let\DiffSpace\relax
    \else
        \ifx\diffarg\{%
        \let\DiffSpace\relax
      \fi\fi\fi\DiffSpace}

\usepackage{tikz,xcolor,hyperref}

\definecolor{lime}{HTML}{A6CE39}
\DeclareRobustCommand{\orcidicon}{\hspace{-1mm}
	\begin{tikzpicture}
	\draw[lime, fill=lime] (0,0) 
	circle [radius=0.16] 
	node[white] {{\fontfamily{qag}\selectfont \tiny \,ID}};
	\draw[white, fill=white] (-0.0525,0.095) 
	circle [radius=0.007];
	\end{tikzpicture}
	\hspace{-3mm}
}

\foreach \x in {A, ..., Z}{\expandafter\xdef\csname orcid\x\endcsname{\noexpand\href{https://orcid.org/\csname orcidauthor\x\endcsname}
			{\noexpand\orcidicon}}
}


\newcommand{\mytitle}{Shedding Infrared Light on QCD Axion and ALP Dark Matter with JWST}

\begin{document}


\title{\mytitle}

\author{Akash Kumar Saha\orcidA{}}
\email{akashks@iisc.ac.in}
\affiliation{Centre for High Energy Physics, Indian Institute of Science, C.\,V.\,Raman Avenue, Bengaluru 560012, India}

\author{Subhadip Bouri\orcidB{}}
\email{subhadipb@iisc.ac.in}
\affiliation{Department of Physics, Indian Institute of Science, C. V. Raman Avenue, Bengaluru 560012, India}
\affiliation{Centre for High Energy Physics, Indian Institute of Science, C.\,V.\,Raman Avenue, Bengaluru 560012, India}

\author{Anirban Das\orcidC{}}
\email{anirbandas.21@protonmail.com}
\affiliation{Theory Division, Saha Institute of Nuclear Physics, 1/AF, Bidhannagar, Kolkata 700064, West Bengal, India}
\affiliation{Homi Bhabha National Institute, Training School Complex, Anushaktinagar, Mumbai
400094, India}

\author{Abhishek Dubey\orcidE{}}
\email{abhishekd1@iisc.ac.in}
\affiliation{Centre for High Energy Physics, Indian Institute of Science, C.\,V.\,Raman Avenue, Bengaluru 560012, India}

\author{Ranjan Laha\orcidD{}}
\email{ranjanlaha@iisc.ac.in}
\affiliation{Centre for High Energy Physics, Indian Institute of Science, C.\,V.\,Raman Avenue, Bengaluru 560012, India}

\date{\today}


\begin{abstract}
James Webb Space Telescope (JWST) has opened up a new chapter in infrared astronomy. Besides the discovery and a deeper understanding of various astrophysical sources, JWST can also uncover the non-gravitational nature of dark matter (DM). If DM is QCD axion or an eV-scale Axion-like particle (ALP), it can decay into two photons in the infrared band. This will produce a distinct line signature in the spectroscopic observations made by JWST. Using the latest NIRSpec IFU spectroscopic observations from JWST, we put the strongest bound on the photon coupling for QCD axion/\,ALP DM in the mass range between 0.47 and 2.55\,eV. In particular, we are able to probe a new mass range for ALP DM between $\sim$ 0.47 eV to 0.78 eV beyond what can be probed by globular cluster observations. We constrain well-motivated and UV complete models of QCD axion and ALP DM, including predictions from some models derived from string theory and/\,or various Grand Unification scenarios. Future JWST observations of DM-rich systems with a better understanding of the astrophysical and instrumental backgrounds can thus enable us to potentially discover QCD axion and ALP DM. The datasets used in this work are available at: \href{https://archive.stsci.edu/doi/resolve/resolve.html?doi=10.17909/3e5f-nv69}{https://archive.stsci.edu/doi/}. 
\end{abstract}

\maketitle

\section{Introduction}
\label{Introduction}

Indirect search programs for DM have received a boost in  recent years from multiple space and ground-based telescopes.  Non-gravitational interaction of DM with Standard Model (SM) particles may cause it to decay or annihilate into known stable particles, like photons, electrons, protons, etc. These processes will dominantly occur in regions with high DM density, such as the Milky Way (MW) DM halo, dwarf galaxies, and other extragalactic halos. The flux of these particles can be detected by various telescopes depending on DM decay/annihilation channels which, in turn, is dictated by the particle physics model of DM\,\cite{Cirelli:2024ssz,Bertone:2016nfn,Strigari:2012acq,Slatyer:2017sev}. 

\begin{figure}[!t]
	\begin{center}
	\includegraphics[width=\columnwidth]{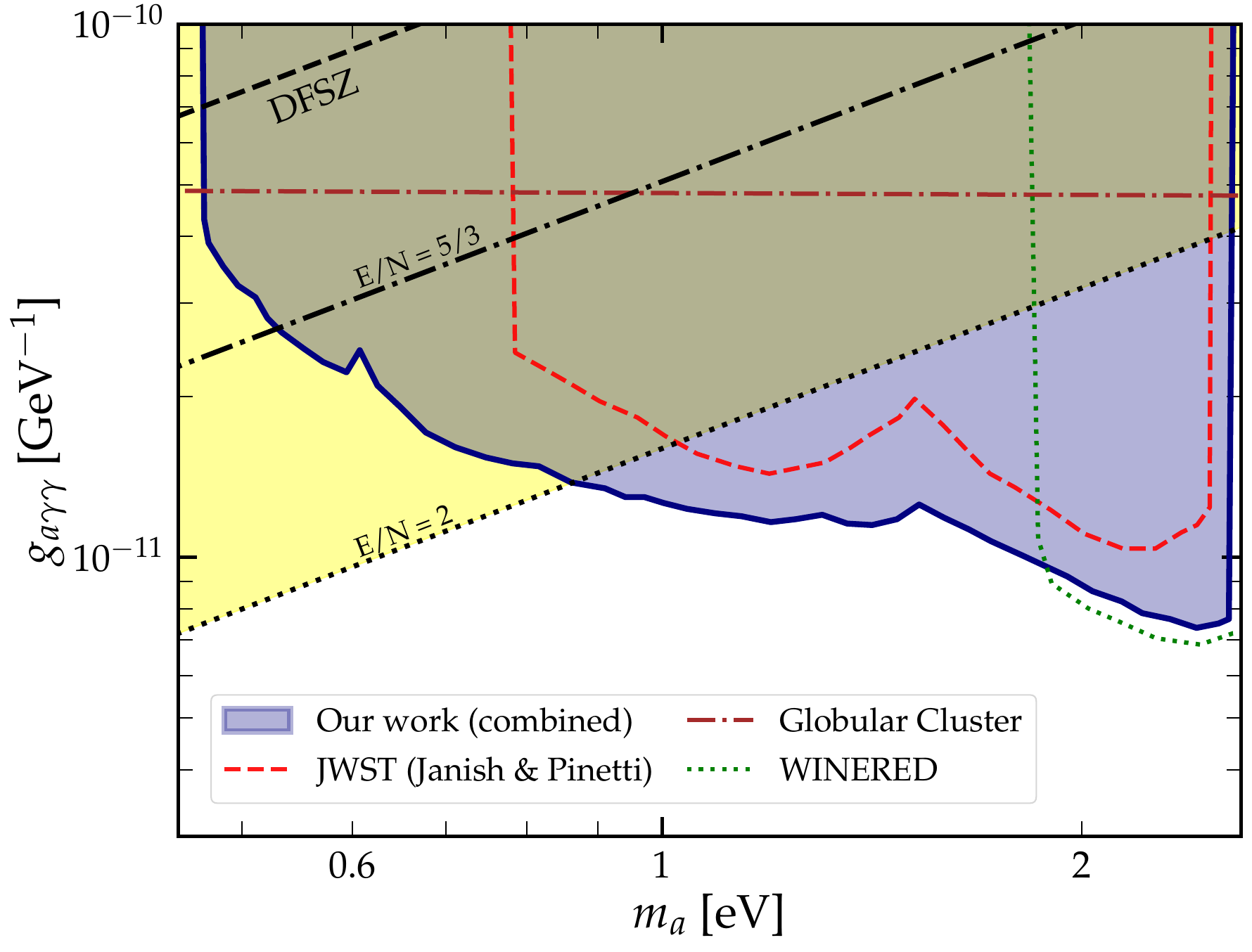}
	\caption{ Schematic of our combined limits on axion-photon coupling, $g_{a\gamma\gamma}$, as a function of the axion mass, $m_a$, obtained using the JWST NIRSpec spectroscopic observations (light blue shaded region). The black dashed line shows the QCD axion parameter space for the DFSZ model. The yellow shaded region shows the `QCD axion' window\,\cite{DiLuzio:2016sbl,DiLuzio:2017pfr,Agrawal:2024ejr}. Other existing limits in the parameter space include JWST observation of GNz11 galaxy (red dashed)\,\cite{Janish:2023kvi}, dwarf galaxy observations by WINERED (green dotted)\,\cite{Yin:2024lla}, and stellar evolution estimates from Globular Cluster (brown dot-dashed)\,\cite{Ayala:2014pea,Dolan:2022kul}. For more precise limits, please refer  to Fig.\,\ref{fig:-ALP}.
    }
	\label{fig:-JWST rough}
    \end{center}
\end{figure}

 Axions are one of the most well-motivated DM candidates. QCD axions were  introduced as a solution to the strong CP problem\,\cite{PhysRevLett.38.1440,PhysRevD.16.1791,PhysRevLett.40.223,PhysRevLett.40.279}. They can be produced non-thermally in the early universe and thus despite its small mass can contribute significantly to the DM density\,\cite{Abbott:1982af,Dine:1982ah,Preskill:1982cy,PhysRevLett.48.1867,PhysRevD.32.1560,Davis:1986xc}. 
QCD axions also arise via string theory compactifications\,\cite{Banks:2003sx,Svrcek:2006yi,Conlon:2006tq, DiLuzio:2020wdo, Agrawal:2024ejr}. Here the axion-photon coupling depends on how the SM gauge group is embedded into a higher dimensional gauge theory. Thus, a discovery of the QCD axion can  constrain string theory models\,\cite{Arvanitaki:2009fg,Gendler:2024adn}. Unlike QCD axions, Axion-like Particles (ALPs) do not couple to QCD and need not follow the linear mass-coupling relation\,\cite{Daido:2017wwb,Daido:2017tbr}.
 
 Axions have a coupling to photons via the Lagrangian, $\mathcal{L_{\rm int}}=-\frac{1}{4}\,g_{a\gamma\gamma} a F_{\mu\nu}\tilde
F_{\mu\nu}$, where $g_{a\gamma\gamma}$ is the axion-photon coupling strength, $a$ is the pseudoscalar axion field\footnote{In this paper, we use `QCD axion' and `ALP' interchangeably.}, and $F_{\mu\nu}$ ($\tilde F_{\mu\nu}$) is the electromagnetic (dual) field strength tensor.  
Axion  particles of mass $m_a$ can decay into two photons, where the mass of the decaying ALP determines the wavelength of the final-state photons. Decaying ALPs have been searched for in gamma-rays\,\cite{Kolb:1988pe,Bernal:2022xyi,Calore:2022pks,Caputo:2022mah,Hoof:2022xbe}, X-ray\,\cite{Porras-Bedmar:2024uql,Higaki:2014zua,DeRocco:2022jyq,Beaufort:2023zuj,Higaki:2014qua,Foster:2021ngm,Dessert:2023vyl,Diamond:2023cto,Candon:2024eah,Nguyen:2023czp}, ultra-violet\,\cite{Todarello:2024qci,Kar:2025ykb}, optical\,\cite{Grin:2006aw,PhysRevLett.129.231301,Regis:2020fhw,Bernal:2022xyi,Nakayama:2022jza, Carenza:2023qxh, Todarello:2023hdk, Wang:2023imi,Dror:2024ibf}, infrared\,\cite{Kohri:2017ljt,Caputo:2020msf,Bessho:2022yyu,Janish:2023kvi,Roy:2023omw,Yin:2024lla}, and radio\,\cite{Caputo:2018vmy,Chan:2021gjl,Sun:2021oqp,PhysRevD.111.023011,Dev:2023ijb,Arza:2019nta,Arza:2021nec}. These searches have yielded stringent constraints on the masses and couplings of ALP DM. Besides decay, ALPs can also convert into photons in the presence of external magnetic field, and this too results in strong limits\,\cite{CAST:2007jps,CAST:2017uph,CAST:2024eil,HESS:2013udx,Meyer:2016wrm,FermiLAT:2016nkz,Meyer:2020vzy,MAGIC:2024arq,Li:2021gxs,Jacobsen:2022swa,Muller:2023vjm,Li:2024zst,Davies:2022wvj,Xiao:2020pra,Wouters:2013hua,Marsh:2017yvc,Reynolds:2019uqt,Reynes:2021bpe,Mondino:2024rif,Calore:2021hhn,Buen-Abad:2020zbd,Dev:2023hax,Ning:2024eky,Dessert:2021bkv,Dessert:2022yqq,Ruz:2024gkl,DeRocco:2022jyq,Beaufort:2023zuj,Dessert:2020lil,Li:2020pcn,Foster:2022fxn,Battye:2023oac}. Upcoming telescope observations hold the promise of probing new regions of the ALP parameter space\,\cite{Thorpe-Morgan:2020rwc,Caputo:2018ljp,Regis:2024znx,Libanore:2024hmq}.

We use the James Webb Space Telescope (JWST), launched in 2021, to probe new parameter space of ALP DM.  
It has state-of-the-art imaging and spectroscopic capabilities in the infrared waveband, and is equipped with four instruments: Near-Infrared Spectrograph (NIRSpec)\,\cite{jakobsen2022near}, Near-Infrared Camera (NIRCam)\,\cite{rieke2022nircam}, Near-Infrared
Imager and Slitless Spectrograph (NIRISS)\,\cite{doyon2023near}, and Mid-Infrared Instrument (MIRI)\,\cite{argyriou2023jwst}. 
We focus on JWST NIRSpec\,\cite{2022A&A...661A..80J} spectroscopic observations of distant galaxies to search for photon line from
ALP DM decay. Ref.\,\cite{Roy:2023omw} calculated future sensitivities on axion-photon coupling from JWST Blank Sky observations. We also note that in Ref.\,\cite{An:2024kls} JWST datasets have been used to search for Dark Photon DM.

\begin{figure}[!t]
	\includegraphics[width=0.5\textwidth]{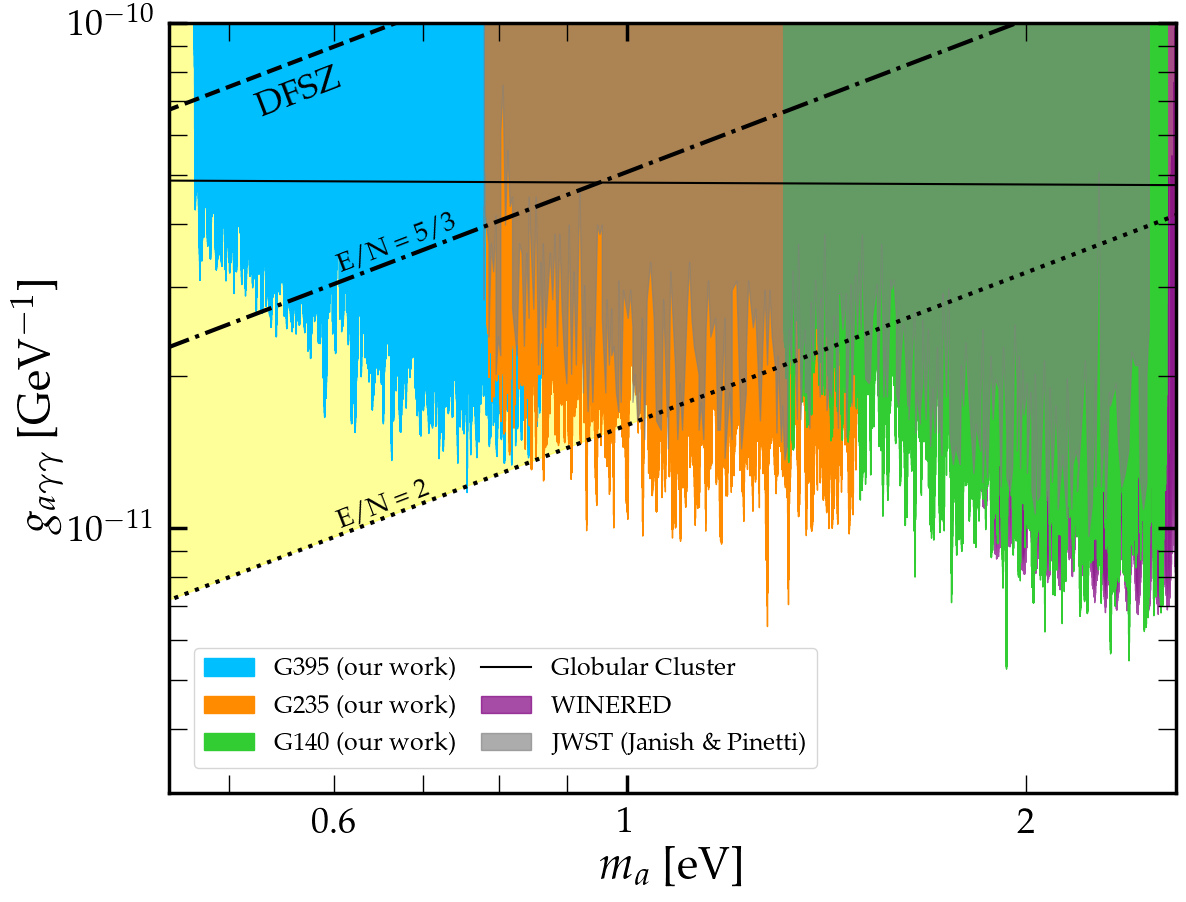}
	\caption{Constraints on ALP-photon coupling of ALP DM using the high-redshift spectroscopic observations by JWST. Our limits are obtained by analyzing the G395\,(M/H) 
 (sky-blue shaded), G235\,(M/H) (orange shaded), and G140\,(M/H)  
 (light green shaded) grating datasets of the NIRSpec instrument. The previous limits include WINERED observation of Leo V and Tucana II dwarf galaxies\,\cite{Yin:2024lla} (purple), JWST observation of GNz11 galaxy\,\cite{Janish:2023kvi} (gray), and stellar observations from Globular Clusters\,\cite{Ayala:2014pea,Dolan:2022kul} (black solid).  }
	\label{fig:-ALP} 
\end{figure}

Our new limit on $g_{a\gamma\gamma}$ as a function $m_a$, shown in Fig.\,\ref{fig:-JWST rough} as a blue shaded region, is the strongest between $\sim\,0.5$ and $2\eV$. Moreover, we exclude a significant part of the QCD axion parameter space denoted by the yellow shaded region. The representative values $E/N=$ 2, 5/3, and 8/3 (DFSZ) are shown by black dotted, black dot-dashed and black dashed lines, respectively, where $E$ is the electromagnetic and  $N$ is the color anomaly coefficients. The exact value of $E/N$ depends on the UV complete model\,\cite{DiLuzio:2016sbl,DiLuzio:2017pfr,Agrawal:2024ejr}.
The previous limits in this mass range include the following: high redshift galaxy GNz11 observations by JWST (red dashed line)\,\cite{Janish:2023kvi}, WINERED observation of dwarf galaxies (green dotted line)\,\cite{Yin:2024lla}, and observation of globular cluster stars (brown dot-dashed line)\,\cite{Ayala:2014pea,Dolan:2022kul}. 
For a more complete limit we refer to Fig.\,\ref{fig:-ALP}.  
We significantly improve upon the existing limits by using a larger dataset and utilizing longer wavelength spectral data from JWST NIRSpec integral field unit (IFU). With this, we are able to probe new regions of the parameter space for decaying QCD axion DM.  


\section{ALP decay signal}
\label{ALP}
An ALP DM of mass $m_a$ will decay into two photons of wavelength, $\lambda_a=4\pi/m_a$.  The decay rate is\,\cite{PhysRevD.54.1} 
\begin{eqnarray}
    \Gamma=\frac{g_{a\gamma\gamma}^2m_a^3}{64 \pi}\,\,,
    \label{rate}
\end{eqnarray}
where $g_{a\gamma\gamma}$ has units of GeV$^{-1}$. The differential flux of photons from the galactic halo ALP DM decay is,
\begin{eqnarray}
\label{flux}
    \Phi_a \equiv \frac{1}{\Delta\Omega}\dfrac{d\phi}{d\lambda} =\frac{m_a}{2}\frac{\Gamma}{4\pi m_a}\frac{dN}{d\lambda}\frac{\mathcal{D}(\ell,b)}{{\Delta\Omega}}\,\,,
\end{eqnarray}
in units of $\surfb$, where $\phi$ is in units of erg cm$^{-2}$s$^{-1}$ and $\Delta\Omega$ is the angular size of the target region. The DM-induced photon line shape (referred to as DM signal) spectrum is
\begin{eqnarray}
    \frac{dN}{d\lambda}=2\times\frac{1}{\sqrt{2\pi w^2}}\,{\rm exp}\left[-\frac{(\lambda-\lambda_a)^2}{2w^2}\right]\,,
   \end{eqnarray}
    with
    \begin{eqnarray}
        w^2=\left (\frac{\Delta \lambda}{2\sqrt{2\, \rm ln2}}\right)^2 + \lambda_a^2\sigma_v^2\,\,.
        \end{eqnarray}
 In the above equation, the first and the second terms take into account the broadening of the photon line due to telescope resolution and DM velocity dispersion in the MW, $\sigma_v \approx 200\,\mathrm{km\,s^{-1}}$\,\cite{Peter:2013aha, Evans:2018bqy}, respectively. The detector resolution is the dominant contribution in our entire wavelength range. The factor $\Delta\lambda$ is related to the spectral resolution of the instrument, $\mathcal{R}$, via $\Delta\lambda=\lambda/\mathcal{R}$ \cite{jwst_disperser}.

In Eq.\,(\ref{flux}), $\mathcal{D}$ is the so-called D-factor that encapsulates the distribution of the ALP DM in the MW along the line-of-sight of the telescope.
Given a DM density profile $\rho (s, \ell, b)$, the D-factor is 
   $ \mathcal{D}(\ell,b)=\int \rho (s, \ell, b) \,ds\,d \Omega\,,$ where $s$ is the line-of-sight distance, and $\ell$, $b$ are the galactic longitude and latitude of the target, respectively. We use the NFW profile for DM density in the MW halo\,\cite{Navarro:1996gj},
\begin{eqnarray}
    \rho(r)=\frac{\rho_s}{\dfrac{r}{r_s}\left(1+\dfrac{r}{r_s}\right)^2}\,\,,
\end{eqnarray}
where $r$ is the distance from the Galactic Center (GC). We take the fiducial values, $\rho_s=0.18\,\mathrm{GeV\,cm^{-3}}$ and $r_s=24$\,kpc\,\cite{Maity:2021umk,sun2025dynamical}. The distance 
$r$ can be converted to the line-of-sight distance, $s$, via, $ r(s,\,b,\,\ell) \,=\, \sqrt{s^2 \,+\, r_\odot^2 \,-\, 2 \, s \, r_\odot \, {\rm cos}\, b \, {\rm cos}\,\ell}$, where $r_\odot=8\,$kpc is the distance of the Sun from the GC.

\section{JWST Data and Analysis}
\label{jwst data}
We use the JWST NIRSpec IFU spectroscopic data publicly available on the MAST portal\,\cite{mast_portal}. For IFU observations, JWST uses a square aperture of size \mbox{3" $\times$ 3"}. For the resulting data, each of the spatial pixel element (`spaxel') is 0.1" $\times$ 0.1". JWST NIRSpec utilizes dispersers as optical elements to spread light into its component wavelengths. These dispersers include three diffraction gratings (G140, G235, and G395) and prism (PRISM). Each of these gratings can provide both medium (`M') ($\mathcal{R}\sim 1000$) and high (`H') ($\mathcal{R}\sim 2700$) resolution spectroscopy.  The filter-disperser combinations used for IFU and their corresponding wavelength ranges can be found in Ref.\,\cite{filter-disperser}. After passing through the dispersers, the incident light is collected in the detector pixels consisting of pixelated HgCdTe detectors\,\cite{pixel_detector}.

Our JWST galaxy datasets\,\footnote{\href{https://archive.stsci.edu/doi/resolve/resolve.html?doi=10.17909/3e5f-nv69}{https://archive.stsci.edu/doi/}} include different source classes like starburst galaxies, elliptical galaxies, quasars, and many others. We use 383 observations (calibration 3) across different JWST IFU gratings (we do not use PRISM due to its low spectral resolution) for this analysis. We only use those JWST target sources for which the decaying ALP DM contribution from the MW halo does not get removed as part of the foreground. In the Supplemental Materials, we elaborate more on the IFU spectral data and various background subtraction methods used by JWST. 

We assume a decaying ALP DM in MW halo and try to search for the photon `line signature' in the spectroscopic observations made by JWST.  Compared to the previous work Ref.\,\cite{Janish:2023kvi}, we utilize a much larger dataset and the data from G395M and G395H gratings that cover longer wavelengths.   The mass ranges that the G395, G235, and G140 gratings probe are $\sim$ 0.47 - 0.86\,eV, 0.78 - 1.5\,eV, 1.31 - 2.56\,eV respectively.  As a result, we not only improve upon  existing limits, but also extend the bounds to lower masses of ALP DM. We are also able to probe a larger region of QCD axion window.

Any JWST spectroscopic observation includes some continuum foreground, background, and instrumental emissions, such as zodiacal light, in-field interstellar medium emission, thermal self-emission, and detector stray light\,\cite{rigby2023dark, 2016SPIE.9904E..0AL,2006SPIE.6265E..3CW, jwst_background}. For a given observation region, these continuum emissions exhibit negligible spectral variations within the DM signal line width. An unexpected `line excess' over the continuum can thus indicate the presence of a decaying ALP DM. 
In some cases, the JWST spectra contain negative flux values occurring due to various instrumental artifacts. We  elaborate on this in the Supplemental Material. 

For the data analysis, we closely follow the methods used by Ref.\,\cite{Janish:2023kvi}. We have used the code provided in Ref.\,\cite{JanishCode} and modified the necessary parts for our analysis. Particularly, we made sure to not include any negative flux values present in the dataset. The details of the data analysis can be found in Supplemental Material.

After finding the continuum fit to the data, we add a DM decay signal. With the added DM signal, we yet again perform a best-fit to the data.  We define a test statistic as  
\begin{equation}
    \chi^2 (g_{a\gamma\gamma},m_a) \equiv \sum_{i} \frac{\left[\Phi_i-\Phi_a(g_{a\gamma\gamma}, m_a) -S(\alpha_i) \right]^2}{\sigma_i^2}\,\,,
    \label{chiSq}
\end{equation}
where $\Phi_i$ is the observed flux in $i^{\rm th}$ wavelength bin, $\sigma_i$ is the corresponding error, and $S(\alpha_i)$ is the continuum spline fit.
We find $\chi_{\rm min}^2$ by minimizing the $\chi^2$ statistic. For a particular $m_a$, we impose a 95\% upper limit on $g_{a\gamma\gamma}$ when $\Delta \chi^2=\chi^2 -\chi^2_{\rm min}=2.71$. We repeat the above analysis for different ALP masses to obtain the final limits on ALP-photon coupling.

\begin{figure}[t]
	\includegraphics[width=\columnwidth]{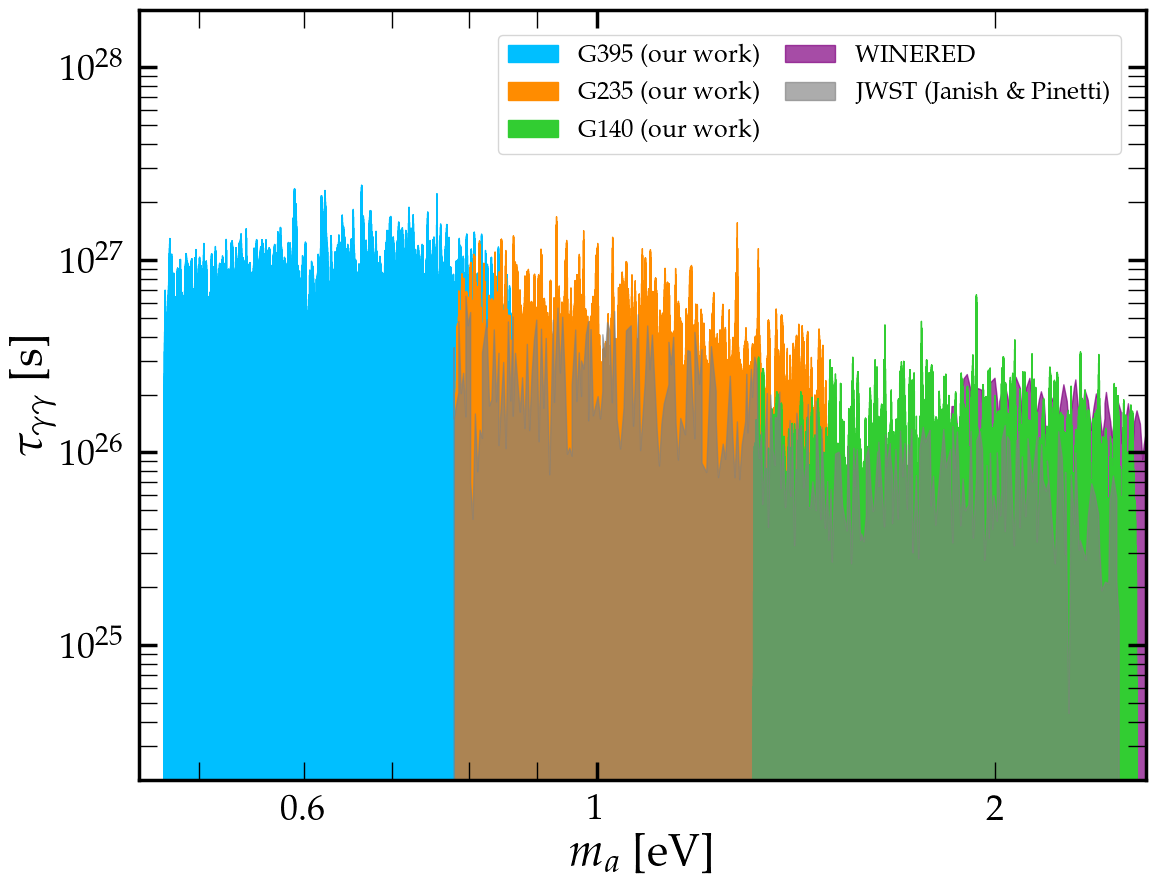}
	\caption{Constraints on the lifetime $\tau_{\gamma\gamma}$ of decaying DM utilizing the JWST IFU spectroscopic observations. Previous limits are same as Fig.\,\ref{fig:-ALP} and are scaled from Refs.\,\cite{Yin:2024lla,Janish:2023kvi}. }
	\label{fig:-ALP_lifetime} 
\end{figure}

\section{Results}
\label{results}

 For each JWST target, we perform the analysis described in Eq.(\ref{chiSq}), and show the combined limit envelope in Fig.\,\ref{fig:-ALP}. Our limits are shown by the sky-blue, orange, and light green shaded regions for G395, G235, and G140 gratings, respectively. For each grating, we have included both medium and high resolution datasets in our analysis.  The previous limits include the combined GNz11 limits, WINERED dwarf observation limits, and Globular Cluster limits, shown by grey shaded, purple shaded, and black solid line, respectively.

Evidently, our analysis rules out a substantial region of the QCD axion parameter space especially using the G395 and G235 grating data. 
 With the JWST G395 grating datasets, we are probing a new ALP DM parameter space, $m_a \sim$ 0.47\,-\,0.78 eV. For G235 and G140 gratings, our limits are stronger than those reported in Ref.\,\cite{Janish:2023kvi} due to the larger dataset used here. 
 Our combined limits obtained using the JWST G140 grating observations are  competitive with the limits from WINERED observation of dwarfs. All limits become weaker at lower masses due to the \(\sim m_a^3\) scaling of the ALP decay width. We found some spurious limits arising due to the fluctuations in the JWST data which we removed manually.

 We also translate our results to a lower limit on the lifetime of a generic DM decaying into two photons as shown in Fig.\,\ref{fig:-ALP_lifetime}.
 For this, Eq.\,\ref{rate} can be modified as, $\Gamma=1/\tau_{\gamma\gamma}$, where $\tau_{\gamma\gamma}$ is the DM lifetime. We show our limits with sky-blue, orange, and light
green shaded regions for the G395, G235, and G140 gratings, respectively, as before. Here too, we are able to probe new regions of parameter space for generic decaying DM.  The previous limits from Refs.\,\cite{Janish:2023kvi, Yin:2024lla} are also scaled appropriately. We note that the Globular Cluster limit is not shown as it depends on the DM model under consideration\,\cite{Boehm:2020wbt}.

\section{Discussion, scope, and conclusion}
\label{discussion}

Our work shows that with the current spectroscopic observations by JWST, we can already probe new parts of the QCD axion window, $m_a \sim$ 0.47\,-\,0.78 eV. Recently, an increasing number of experimental efforts are trying to probe QCD axion at lower masses\,\cite{Ouellet:2018beu,Salemi:2021gck,Liu:2018icu,Pandey:2024dcd,2010PhRvL.104d1301A,ADMX:2018gho,ADMX:2019uok,ADMX:2021nhd,ADMX:2024xbv,ADMX:2018ogs,ADMX:2021mio,Crisosto:2019fcj,Stern:2016bbw,Nagano:2019rbw,Lawson:2019brd,Ahyoune:2023gfw,Devlin:2021fpq,BREAD:2021tpx,Aja:2022csb,Lee:2020cfj,CAPP:2020utb,Yoon:2022gzp,Kim:2022hmg,Yi:2022fmn,Yang:2023yry,Kim:2023vpo,CAPP:2024dtx,Adair:2022rtw,DeMiguel:2023nmz,Obata:2018vvr,Oshima:2023csb,Heinze:2024bdc,Fan:2024mhm,Grenet:2021vbb,Beurthey:2020yuq,Garcia:2024xzc,HAYSTAC:2024jch,HAYSTAC:2023cam,Heinze:2023nfb,McAllister:2017lkb,Quiskamp:2022pks,Quiskamp:2023ehr,Quiskamp:2024oet,Alesini:2019ajt,Alesini:2020vny,Alesini:2022lnp,QUAX:2023gop,QUAX:2024fut,CAST:2020rlf,Ahyoune:2024klt,Wuensch:1989sa,PhysRevLett.59.839,Gramolin:2020ict,Sulai:2023zqw,Arza:2021ekq,Friel:2024shg,TASEH:2022vvu,Hagmann:1996qd,PhysRevD.42.1297,Thomson:2019aht,Thomson:2023moc}. Probing QCD axion DM in sub-eV to eV mass range has proven to be difficult due to technical limitations, though some future proposals are on the horizon\,\cite{Baryakhtar:2018doz, Batllori:2023gwy}. Hence, JWST's reach for heavier QCD axions is an important breakthrough. With more JWST spectroscopic observations at higher wavelengths, these limits are expected to extend even deeper into the QCD axion mass window.

We use JWST spectroscopic observations to put the strongest constraints on QCD axion as well as decaying ALP DM. In principle, the line of sight to any extragalactic astrophysical target includes three different decaying DM components: MW, extragalactic, and host halo. Unlike the MW component, the extragalactic DM spectra is expected to be spectrally smooth covering a wide range of wavelengths. The host halo spectra is a line feature similar to the MW case, shifted to shorter wavelengths due to redshifting of the signal. The datasets used in this work can also be used to constrain heavier ALP DM for the corresponding gratings. But given the distance of these sources used here, the fluxes from extragalactic and host halo ALP DM are sub-dominant compared to the MW DM signal; the resulting limits are also weaker. A detailed analysis of extragalactic and host halo contributions with JWST observations will be an interesting avenue.

The spectroscopic observations by JWST contain many known molecular lines. These lines carry important information about the source classes that JWST is looking at. We have not taken into account the `known' source lines for the DM search. An analysis of the molecular lines along with the exotic DM line signal will result in a more robust limit, and will also be crucial for discovering any possible DM signal at these wavelengths.

Dwarf galaxies are also known to be excellent targets for DM search given their higher DM density and lack of astrophysical backgrounds\,\cite{Yin:2024lla, Regis:2020fhw, Todarello:2023hdk, Todarello:2024qci}. For JWST, dwarfs can provide with cleaner datasets to search for ALP DM. To the best of our knowledge, no JWST dwarf observation datasets have been made public yet. With dedicated observations of dwarf galaxies by JWST, the constraints on ALP DM are expected to improve significantly.

The limits shown in this work also depend on the JWST observation time. The limits improve as $\sqrt{t_{\rm obs}}$, where $t_{\rm obs}$ is the observation time. Future datasets with longer observation times are expected to improve the bounds on QCD axion DM.


Moreover, the current searches are limited by the imperfect modeling of the foreground and background emissions that can easily conceal a weak DM signal. Longer observation time and a better understanding of the astrophysical backgrounds will further improve the limits on ALP parameter space. With better resolution and background subtraction, other upcoming infrared telescopes like PRIMA\,\cite{moullet2023prima} can be instrumental in the quest for eV-scale decaying DM. Given the huge existing and upcoming datasets from different telescopes, our search can thus be extended to probe a wide range of ALP DM masses.\\
{\bf \em Note added.}
Ref.\,\cite{Pinetti:2025owq} appeared on arXiv one day before we submitted the article to the pre-print server.

\section*{Acknowledgements} 
All of the data presented in this paper are obtained from the Mikulski Archive for Space Telescopes (MAST). We especially thank Tony Keyes and Diane Karakla from \href{https://stsci.service-now.com/jwst?id=jwst_index}{JWST NIRSpec help desk} for providing detailed information and clarifications regarding JWST NIRSpec observations. We also thank Judhajeet Basu, Sulagna Bhattacharya, Aryaman Bhutani, Debajit Bose, Andrea Caputo, Debtosh Chowdhury, Marco Cirelli, Deep Jyoti Das, Amol Dighe, Ujjal Kumar Dey, Jaya Doliya, Raghuveer Garani, Joachim Kopp, Tarak Nath Maity, Ranjini Mondol, Rachana, Nirmal Raj, Sourov Roy, Sougata Sarkar, Tracy Slatyer, and Marco Taoso for useful comments and discussions. The previous limits shown in this work are obtained from Ref.\,\cite{AxionLimits}. A.K.S. acknowledges the Ministry of Human Resource Development, Government of India, for financial support via the Prime Ministers’ Research Fellowship (PMRF). S.B. acknowledges the Council of Scientific and Industrial Research (CSIR), Government of India, for supporting his research under the CSIR Junior/Senior Research Fellowship program through grant no. 09/0079(15488)/2022-EMR-I. A.D. acknowledges the financial support provided by the Ministry of Education (MoE), Government of India, under the Senior Research Fellowship (SRF). A. Das was supported by the Government of India DAE project No. RSI 4001. R.L.\,\,acknowledges financial support from
the institute start-up funds and ISRO-IISc STC for the
grant no. ISTC/PHY/RL/499. 

The first two authors contributed equally to this
work.

\bibliographystyle{JHEP}
\bibliography{ref.bib}

\clearpage
\newpage

\clearpage
\newpage
\maketitle
\onecolumngrid
\begin{center}
\textbf{\large \mytitle}

\vspace{0.08in}
{ \it \large Supplemental Material}\\ 
\vspace{0.12in}

{Akash Kumar Saha, Subhadip Bouri, Anirban Das, Abhishek Dubey and Ranjan Laha}
\end{center}
\onecolumngrid
 \setcounter{figure}{0}
\setcounter{section}{0}
 \makeatletter
 \renewcommand{\theequation}{S\arabic{equation}}
 \renewcommand{\thefigure}{S\arabic{figure}}

 




\section{JWST fits file content} For our analysis we have used the `\texttt{.x1d}' JWST data files available in Ref.\,\cite{mast_portal}. We note that JWST NIRSpec has a unique ability of acquiring `cube data'. This means that JWST can get both the spectral and 2D spatial information of the incident photon on the sky simultaneously. In the `\texttt{.x1d}' files, the source spectrum as a function of wavelength is obtained by integrating spatial information. As the MW DM density do not substantially change given the small size of the target, we do not use the spatial spectral information in our work and use only the integrated spectral information from `\texttt{.x1d}' files.

The source type for the NIRSpec observation (SRCTYAPT) can be EXTENDED, POINT or UNKNOWN. This distinction depends on how much of the FoV is covered by the source. The headers of the files include WAVELENGTH, FLUX, SURF BRIGHT, BACKGROUND etc. The wavelength (WAVELENGTH) is provided in units of $\mu$m. Source flux (FLUX) is provided in units of Jy, whereas surface brightness (SURF BRIGHT) is in units of MJy/sr. For EXTENDED observations, due to the extended nature of the source, JWST IFU pixels can provide per steradian observation of the flux. 
The same is done for UNKNOWN sources. For POINT sources, the total flux (FLUX) is provided as the source doesn't cover that many pixels. 
However, note that for point source observations, the surface brightness column is not empty. The column entries are obtained from flux, where the angular size is to be taken for only those pixels that were covered by the source. More details can be found in the `Output' section of Ref.\,\cite{Header}.



\section{JWST NIRSpec IFU Background estimation}

In order to remove background, JWST NIRSpec IFU uses two main observation techniques: dither and nod. In dither, consecutive observations are taken with small offsets to help mitigate instrumental and cosmic ray backgrounds. On the other hand, larger offsets are considered in different nods to estimate the source foreground/ background, which then can be subtracted from the main observation. IFU dither values are either NONE or 4-POINT-DITHER. For nod the configurations are NONE, 2-POINT-NOD, and 4-POINT-NOD.

For EXTENDED case, the dedicated background subtraction is optional. For point sources, usually 2-point and 4-point nod patterns or annular background subtraction is performed to remove instrument noise and other foregrounds (including emission from the MW). 
As discussed earlier, for EXTENDED and UNKNOWN sources, SURF BRIGHT header contains the source flux. The background (from MW foreground and instrumental noise) is captured in the BACKGROUND header. If no dedicated background analysis is requested for a particular observation (which is the case for the EXTENDED and UNKNOWN sources), the JWST pipeline implements a default background estimation. For each slice in the cube data, the pipeline performs a sigma clipping, i.e., if the background flux is constant across the FoV, then the source flux will be clipped at a certain threshold where the photon count from the source is very high. The remaining unclipped flux will be summed and considered as the background. For POINT sources, only a small part of the FoV is occupied by the source. So the default clipping can identify the source flux over the constant background. However, for EXTENDED and UNKNOWN sources, most of the pixels get contributions from the source. Here a simple clipping will thus just identify the large fluctuations of the source,  but cannot detect the full source flux. JWST pipeline performs the default sigma clipping for POINT, EXTENDED, and UNKNOWN sources. Therefore for EXTENDED and UNKNOWN sources JWST pipeline cannot remove the background unless a dedicated background estimation is performed. Thus the BACKGROUND contains both source flux and MW foreground. 

We want to emphasize that the search for decaying ALP DM signal from the MW within the spectra of distant targets, the foreground must be contained in the data. Given the sigma clipping method discussed earlier, we note that the DM signal is spatially uniform within the FoV of NIRSpec IFU, and the clipping method will not be able to remove the DM signal from the data. If the dataset provided by JWST does a proper foreground removal, then the MW DM decay signal can also get removed. Thus it is extremely important to make sure that the source spectra have foreground contributions. For our analysis, we use the EXTENDED and UNKNOWN source types, where a dedicated foreground/ background removal is not performed. 

\begin{figure*}[t!]
	\centering
	\includegraphics[scale=0.68]{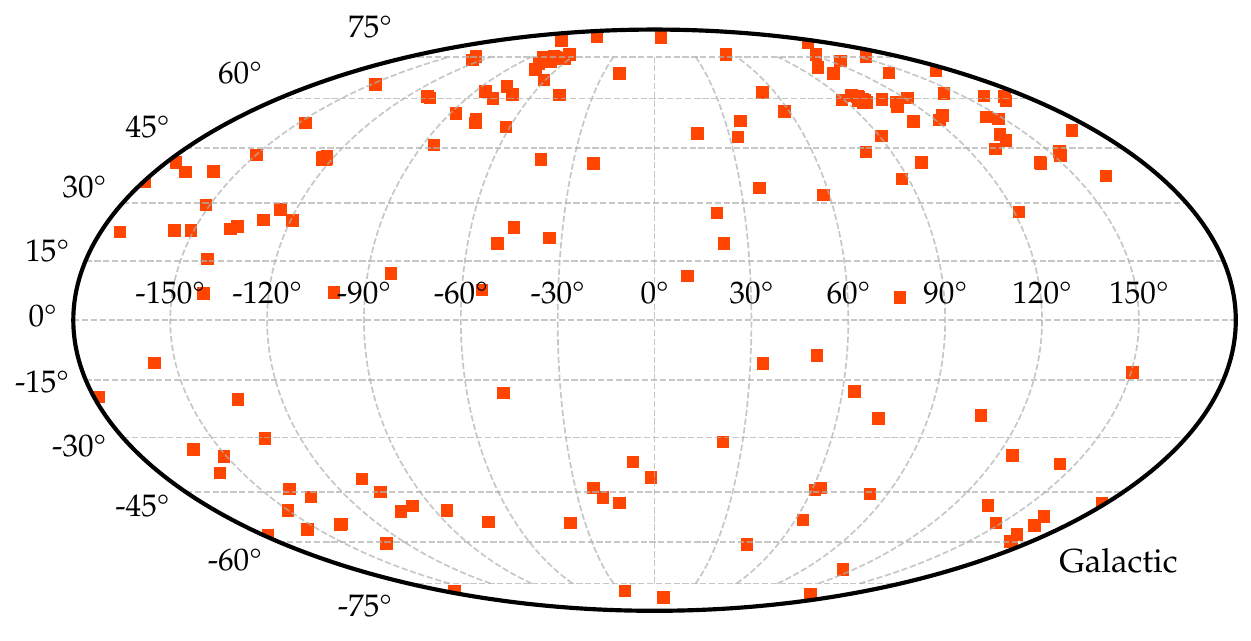}
	\caption{Locations of JWST NIRSpec IFU observations used in this analysis are shown with red squares on a sphere using the galactic coordinate system. The datasets used in this work are available at: \href{https://archive.stsci.edu/doi/resolve/resolve.html?doi=10.17909/3e5f-nv69}{https://archive.stsci.edu/doi/}.
    }\label{fig:source_loc}
\end{figure*}

In Fig.\ref{fig:source_loc}, we show the positions of the JWST observations used in this analysis. Notably, there is no observation within galactic latitude $\pm\,5^\circ$ likely to avoid the galactic plane background contamination. There are 17 observations in the sky map within $\pm\, 15^\circ$. Future GC observations can put stronger constraints in the $g_{a\gamma \gamma}-m_a$ parameter space, as the expected DM decay signal will be enhanced by the increased D-factor. 
Our analysis utilizes 383 datasets from 267 distinct target sources.

\begin{figure}
	\begin{center}
	\includegraphics[width=\columnwidth]{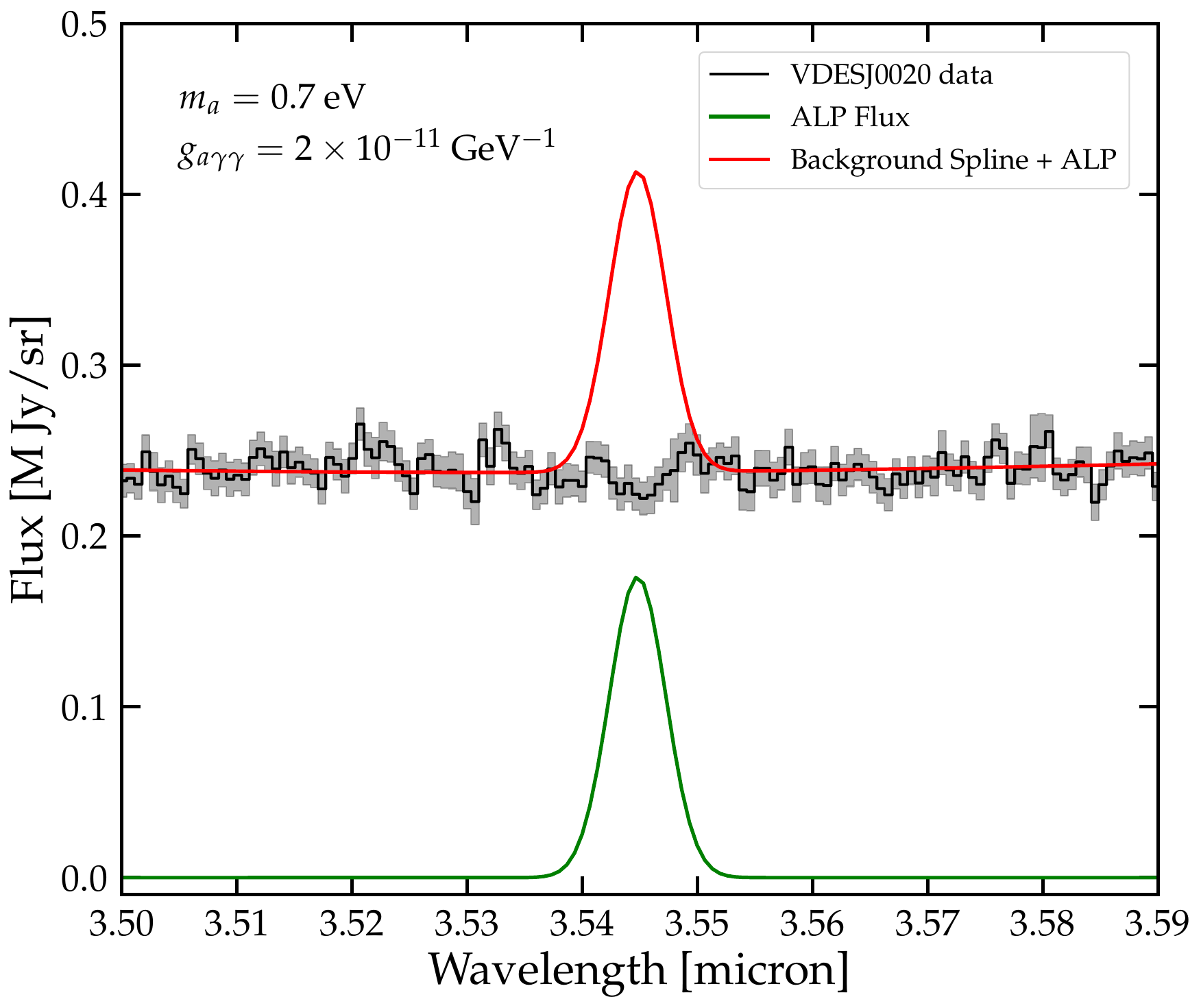}
	\caption{Comparison of the spectroscopic data from JWST observation of VDESJ0020 galaxy and the galactic decaying ALP DM flux for $m_a=0.7$ eV and $g_{a\gamma\gamma}=2\times10^{-11}$GeV$^{-1}$. The observational data is shown by the black solid line and the grey shaded region shows the uncertainty in flux measurement. The galactic ALP decay spectra is shown by the green solid line. The red solid line shows the combined flux of the continuum best-fit background model and ALP DM signal.   }
	\label{fig:-JWST data} 
     \end{center}
\end{figure}

\section{Data analysis}
For a given JWST observation, to find the underlying continuum emission, we employ a cubic spline fitting method\,\cite{spline1,spline2,spline3,spline4}.
 The cubic spline is a piecewise 3rd order polynomial function parametrized by a set of uniformly distributed anchor points ($\alpha_i$) where the polynomial pieces meet. We choose positive chunks of spectra where at least 500 continuous bins do not have any negative flux value. We do not choose smaller positive chunks as those may be substantially affected by instrumental artifacts. Given the large size of the dataset used in this work, we can constrain the whole ALP DM mass range $\sim$ 0.47- 2.55\,eV without any gap. After choosing the spectrum segments, following Ref.\,\cite{Janish:2023kvi}, we do not fit the whole spectrum at a time; instead, for each DM mass $m_{a}$, we focus on local segments centered around decay wavelength $\lambda_{a}$, and the width of the sub-region is set to 150 times the Full Width at Half Maximum (FWHM) of the DM signal line width. While performing the analysis, we make sure that the width of the sub-region of the spectra associated with a particular $m_{a}$ is entirely contained  within the positive chunks of the flux values.

The continuum spline model, $S(\alpha_i)$, is a function of anchor points at fixed wavelengths $\lambda_i$ such that it matches with the observed data only at the anchor points $\alpha_i$. We have set the fiducial number of anchor points to 5. The residuals between the observed data and the spline model are calculated, and the residuals are minimized to find the best-fit spline parameters (anchor points). Next, the residuals are weighted by the errors.   The standard deviation of weighted residuals in each data chunk is used to identify the outliers in the data points. The data points with residuals exceeding a threshold value (set as a clipping factor multiplied by the standard deviation) are excluded from further analysis. The errors are then scaled by the standard deviations of the residuals from the unclipped data. In our case, the resulting scaled errors are $\sim$ 2-3 times higher than the original values. After adding the DM line signal, the fitting procedure described above is again repeated.

We also note that the JWST datasets are provided in solar barycentric frame. We implement Doppler corrections to JWST data in order to transform them to the galactocentric frame before doing our analysis. For this, depending on the source position, we first evaluate the unit vector from Solar position to the source position in the galactocentric frame. We then project the Solar velocity along this unit vector to calculate the corresponding Doppler shift of the spectra in the galactocentric frame.

We have checked our limits for the cases with 100 and 200 times the DM signal  FWHM. Our results do not change by more than 12\%. We have also checked our limits with a 4th order spline fit to the dataset. The limits presented here are robust against the order of the spline fit.

Fig.\,\ref{fig:-JWST data} shows the comparison of the ALP decay signal along with the JWST observation of the VDESJ0020 galaxy (for the G395H grating). For DM decay signal (green solid line) ALP DM with $m_a =$ 0.7 eV and $g_{a\gamma\gamma}=2\times10^{-11}$ GeV$^{-1}$ is considered. The data and the error bars are shown by the black solid line and gray shaded band, respectively. We add the best-fit continuum with the DM signal to obtain the total flux, shown by the red solid line. In this case, the resulting ALP DM decay signal shows an excess over the observation made by JWST, and thus can be excluded.

\section{JWST MOS observation for DM search}

JWST NIRSpec conducts multi-object spectroscopy (MOS) observations using the the micro-shutter assembly (MSA)\,\cite{MSA}. With a total 3.6' $\times$ 3.4' field of view (FoV), MOS provides high-resolution spectroscopy in the wavelength range $\sim$ 0.6–5.3 $\mu$m. MSA has 4 quadrants with a total of $\sim$ 250,000 slitlets with their individual shutters, each having an area 0.20" $\times$ 0.46". The key advantage of MSA relies on the fact that in a single observation, JWST can collect data from multiple sources. Many of the JWST MOS observations can be found in Ref.\,\cite{mast_portal}. 

However, the MSA observation uses a multi-slitlet nod pattern for background removal.  In this method, the target source is placed in different slitlets one after another and data is collected. Assuming these exposures contain a constant foreground/ background, JWST pipeline takes the average of the exposures without the source, and then subtracts it from the source flux. This can be seen in Fig.\,1 of Ref.\,\cite{Nod}. 

Given the MW DM distribution does not change appreciably within the FoV of MOS observations, the nodding performed in MOS will thus subtract any DM decay signal present as the foreground in the data. Similarly extragalactic DM decay signals will also remain constant over the MOS slitlets. Thus, the MOS data cannot be used for MW or extragalctic decaying ALP DM analysis. We note that, the host halo DM distribution can significantly change over the MOS FoV depending on the distance of the source under consideration. Therefore one can use MOS observations to probe decaying QCD axion DM in the host halo, though given the large distance to the halo, the limits are expected to be weak.

\begin{figure}[t!]
	\begin{center}
	\includegraphics[width=\columnwidth]{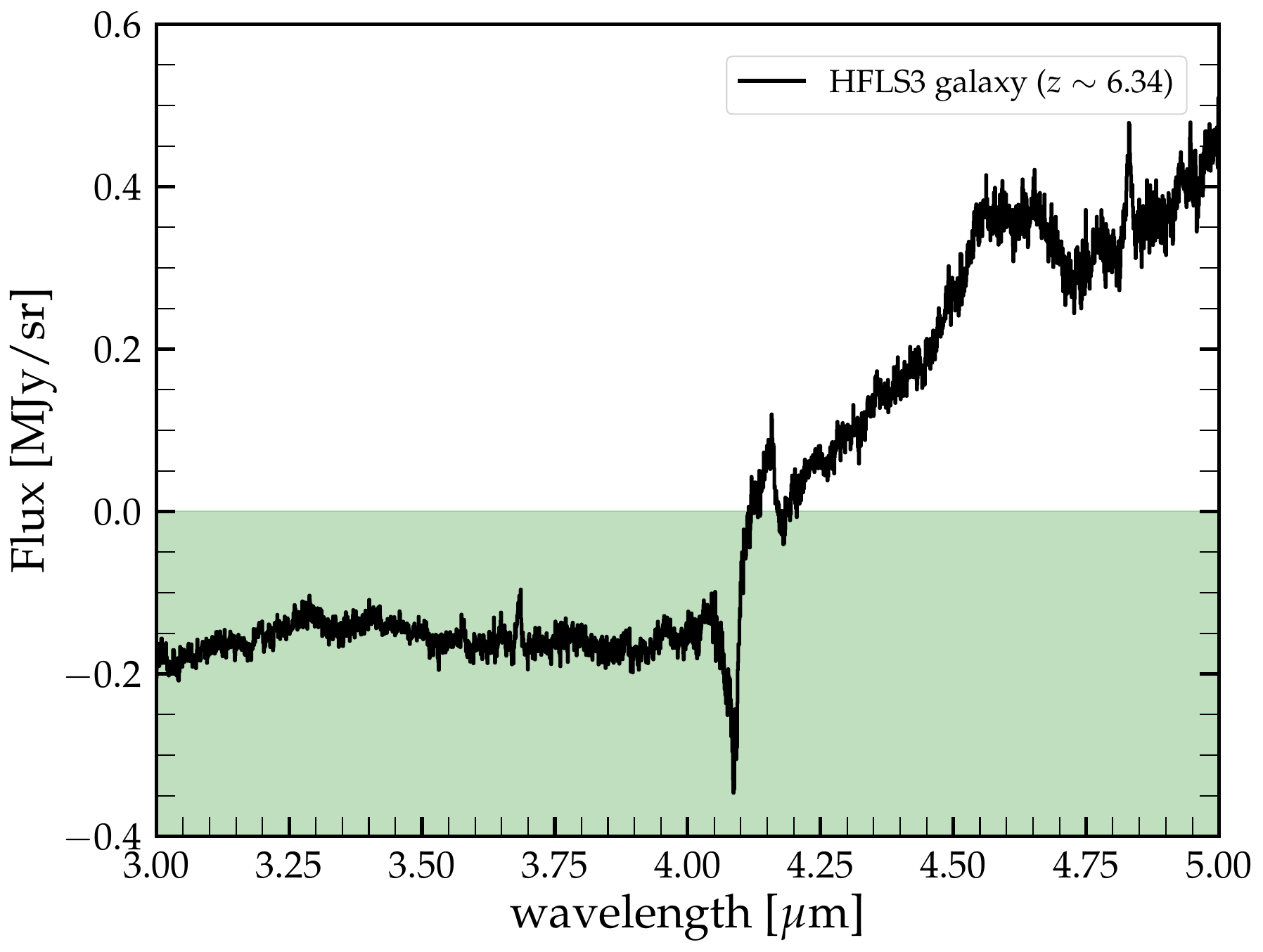}
	\caption{Spectroscopic data from JWST observation of HFLS3 galaxy (observation ID: jw01264-o012-t009-nirspec-g395h-f290lp). Here the header BACKGROUND is plotted as a function of WAVELENGTH (shown in black). The green shaded region show the regions where the observed  background flux is negative.}
	\label{fig:-Negative} 
     \end{center}
\end{figure}
\section{Regarding negative flux values in JWST NIRSPEC IFU observations}
We see that some of the JWST 1D spectra datasets contain negative values. The negative count rates are caused by over-subtraction, which may happen for various reasons. One possibility is the background exposure is slightly higher than the science exposure, leading to an over-subtraction of the background. Another possibility is a mismatch in the dark current (leakage) between the dark reference file and the science exposure due to tiny temperature variations\,\cite{jwst_dark}. 

Several methods can be implemented to address these issues\,\footnote{\mbox{We thank the experts from \href{https://stsci.service-now.com/jwst?id=jwst_index}{JWST NIRSpec help desk} for clarifications.}}. One way is 1/$f$\,-\,noise correction, which can help bring the background level to positive values. Additionally, using a better version of dark files can help mitigate this over-subtraction issue. JWST team has informed us that they will soon release new dark files, which will account for more accurate instrumental artifacts, such as light leaks or temperature effects. 

In Fig.\,\ref{fig:-Negative}, we show an example of such negative flux.We show the NIRSpec IFU observation of the HFLS3 galaxy by black solid line. We note that below $\sim$ 4.25 $\mu$m, the data contains substantial negative values. We avoid such negative flux in our DM analysis as explained in the paper. With a better understanding of instrumental effects in JWST, our limits on ALP DM can be made more robust in future.


\end{document}